\documentclass[11pt,a4paper,twocolumn]{NSF-article}
\usepackage[T1]{fontenc}
\usepackage[utf8]{inputenc}
\usepackage{multicol}
\usepackage{graphicx}
\usepackage{dcolumn}
\usepackage{bm}
\usepackage{upgreek}
\usepackage{fancyhdr}
\usepackage{wrapfig,amsmath,amssymb,enumerate }
\usepackage{cite}
\usepackage{wrapfig}
\usepackage[final]{pdfpages}
\usepackage{titlesec}
\usepackage{abstract}
\usepackage{xcolor}
\titleformat*{\section}{\Large\bfseries}
\titleformat*{\subsection}{\normalsize\bfseries}
\titleformat*{\subsubsection}{\large\bfseries}
\titleformat*{\paragraph}{\large\bfseries}
\titleformat*{\subparagraph}{\large\bfseries}
\usepackage{times}
\usepackage{physics}
\usepackage[labelfont=bf]{caption}

\renewcommand{\abstractname}{}    
\usepackage[obeyspaces]{url}
\usepackage[symbol]{footmisc}

\begin{document}

\twocolumn[
\Large \textbf{Probing non-exponential decay in Floquet-Bloch bands \vspace{2mm}\\}
\noindent \normalsize Alec Cao, Cora J.\ Fujiwara, Roshan Sajjad, Ethan Q.\ Simmons, Eva Lindroth, and David Weld$^*$
\vspace{6mm}]
\footnotetext{$^*$ Corresponding author: {\bf David Weld, Alec Cao, Cora J.\ Fujiwara, Roshan Sajjad, Ethan Q.\ Simmons}: Physics Department, University of California, Santa Barbara, CA USA. {\bf Eva Lindroth}: Department of Physics, Stockholm University, AlbaNova University Center, Stockholm, Sweden.}

\noindent
\textbf{Abstract:} 
Exponential decay laws describe systems ranging from unstable nuclei to fluorescent molecules, in which the probability of jumping to a lower-energy state in any given time interval is static and history-independent.   These decays, involving only a metastable state and fluctuations of the quantum vacuum, are the most fundamental nonequilibrium process, and provide a microscopic model for the origins of irreversibility.  Despite the fact that the apparently universal exponential decay law has been precisely tested in a  variety of physical systems~\cite{Norman_expdecaytests}, it is a surprising truth that quantum mechanics {\em requires} that spontaneous decay processes have non-exponential time dependence at both very short and very long times~\cite{sakurai,Greenland_seeking}. Cold-atom experiments both classic~\cite{raizen_quantumzeno} and recent~\cite{dominik_spontemission} have proven to be powerful probes of fundamental decay processes;
in this paper, we propose the use of Bose condensates in Floquet-Bloch bands as a probe of long-time non-exponential decay in single isolated emitters. We identify a range of  parameters that should enable observation of long-time deviations, and  experimentally demonstrate a key element of the scheme: tunable  decay between quasienergy bands in a driven optical lattice.


\section{Introduction}

Given the ubiquity of exponential decay, it is surprising that quantum mechanics requires that decay processes to a continuum with a ground state exhibit non-exponential long-time dynamics~\cite{khalfin,winter,Fonda_1978,Greenland_seeking,sakurai,knight}. Classic experiments on the subject include negative results from studies of $^{56}$Mn nuclear decay tests~\cite{Norman_expdecaytests} and an indirect observation claimed in investigations of $^{8}$Be scattering phase shifts~\cite{indirectevidencenonexpdecay}. More recently, a variety of physical systems ranging from integrated photonics~\cite{integratedphotonics_nonexp}  to Feshbach molecules~\cite{feshbach_li6} have emerged as platforms for the exploration of non-exponential decay. Extensive theoretical work has been directed toward non-exponential decay of autoionizing resonances in atomic systems~\cite{autoionize_nicolaides,autoionize_druger,nonexp-atomic-decay-calc} and laser-induced ionization effects~\cite{lewenstein-photodetachment,nonexp-tunneling-ionization}, though this remains at the frontier of experimental feasibility.

Negative ions are often considered in this context, in part due to their simple structure: there is usually only one bound state and a few resonances
which simplify the the study of laser-induced negative ion photodetachment~\cite{lewenstein-photodetachment}. Another reason~\cite{Greenland_seeking,nonexp-atomic-decay-calc} is the possibility of finding broad resonances decaying with a very small energy release which, as  discussed below, should result in a deviation at an earlier time when more is left of the parent. On the experimental side however, negative ions also pose certain difficulties, especially due to the low target densities available. To our knowledge, no experiments on non-exponential decay in negative ions have been reported.

\begin{figure}[t]
\centering
\vspace{-.2in}
\includegraphics[width=0.8\columnwidth]{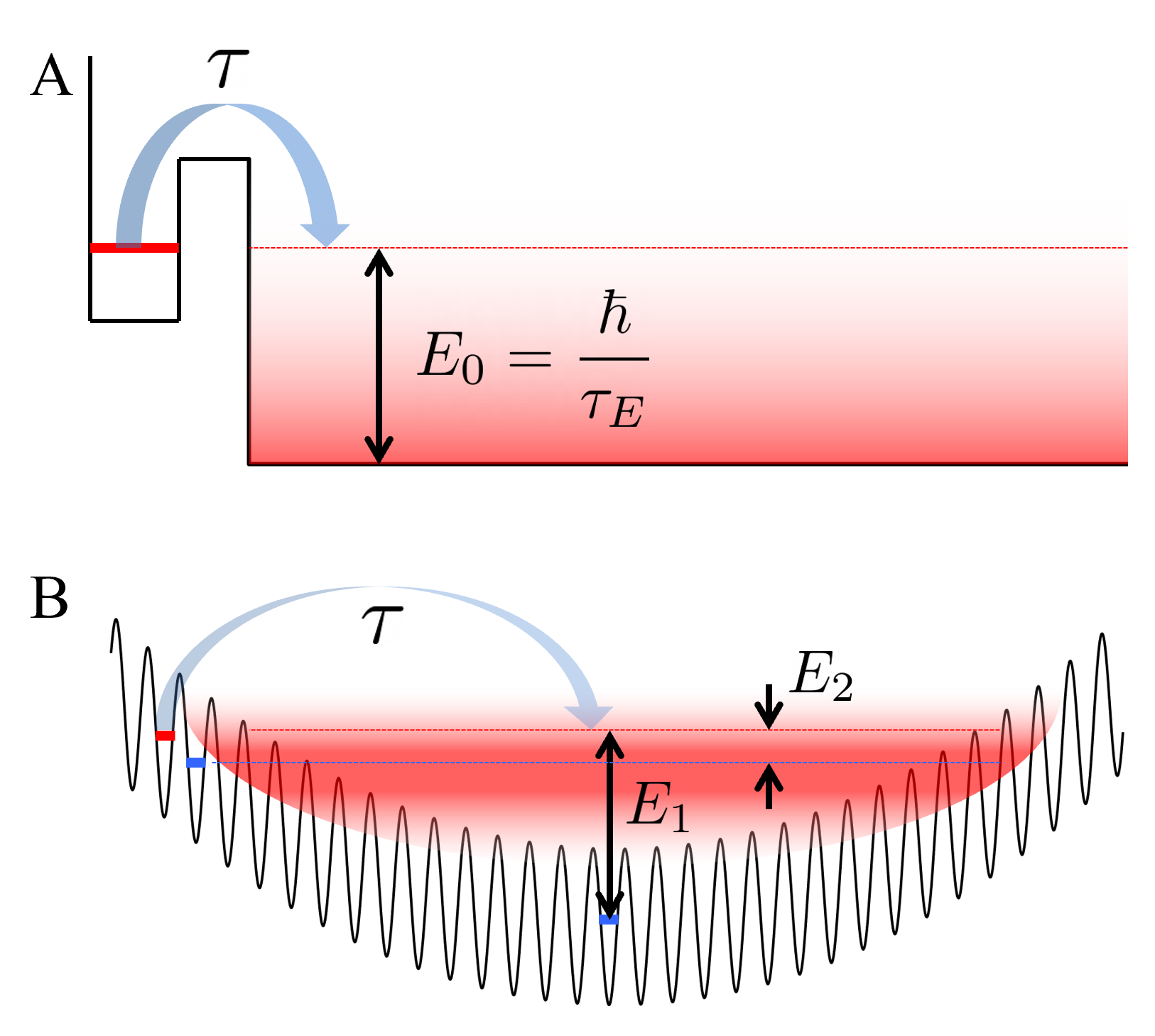}
\caption{A) Schematic of a potential in which non-exponential decay is expected. $\tau$ is the decay time of the exponential part of the tunneling process, and $\tau_E=\hbar/E_0$ is the timescale associated with the energy of the decay product. B) Schematic of proposed optical lattice experiment probing non-exponential decay. $E_1$ and $E_2$ are different possible characterizations of the decay product energy.}
\label{fig:nonexpschematic}
\end{figure}

In a very different physical context, cold atoms in optical lattices can also serve as a probe of decay dynamics~\cite{holthaus_BlochZener}, as shown for example in two seminal experiments. The quantum Zeno effect was first detected using cold sodium atoms in an accelerated optical lattice~\cite{raizen_quantumzeno}; more recently, non-Markovian long-time dynamics were observed in an optically dense ensemble of lattice-trapped atoms driven by an applied microwave field~\cite{dominik_spontemission,dominik_weisskopf}. These results demonstrate the promise of degenerate gases in optical lattices for observing long time modifications to memoryless exponential decay in an ensemble of single emitters.

Here we propose the use of ultracold non-interacting $^7$Li Bloch oscillating in a tilted modulated optical lattice to directly observe long-time non-exponential interband decay. A schematic of the proposed setup and its relationship to an idealized decay process is presented in Figure~\ref{fig:nonexpschematic}. While the proposed experiments can in principle be performed in unmodulated lattices (in close analogy to Ref.~\cite{raizen_quantumzeno} and to pioneering experiments in optical lattice  St\"uckelberg interferometry~\cite{weitz}), we will show that signatures of non-exponential long-time evolution can be greatly enhanced using recently developed tools of Floquet engineering for modification and mapping of band structure~\cite{warpdrivePRL,RSBO-PRL}.

The proposed platform for the exploration of non-exponential decay has several unique advantages. Most important is the extreme tunability afforded by the use of flexible Floquet engineering techniques. Another key advantage, arising from the choice of atomic species, is the presence of broadly Feshbach-tunable interactions in $^7$Li. In this work we emphasize the ability to access the single-emitter regime by tuning the scattering length to zero. However, the ability to work at arbitrary scattering length may also enable future systematic study of the effects of interactions on spontaneous decay.

In Section \ref{sec:theory} of this manuscript, we review a heuristic explanation for non-exponential decay based on a simple analysis of the survival probability and the Breit-Wigner energy distribution. We present numerical calculations of the emergence of non-exponential behavior as a result of imposing a lowest energy bound, revealing decay rate and decay energy as key parameters for experimental observation. In Section \ref{sec:experiment}, we discuss the details and feasibility of the proposed experiment. In particular we experimentally demonstrate the use of Floquet engineering to engineer the band gap and tune the decay rate, a key step on the path to realization of long-time non-exponential decay of an isolated emitter. Section 4 offers conclusions and outlook.

\section{Origins of non-exponential decay\label{sec:theory}}

We begin by recalling a heuristic argument for non-exponential decay which makes no reference to the particular form of the unstable state or decay mechanism~\cite{Fonda_1978}. Given some initial state $\ket{\psi_0}$ with Hamiltonian $H$, the survival or undecayed amplitude $A(t)$ can be calculated as the overlap of the initial state with the time evolved state $\exp(-i H t / \hbar) \ket{\psi_0}$. For a continuous spectrum, the time evolved state can be expanded over the complete set of energy eigenstates $\ket{\phi_E}$ as 
\begin{equation}
    e^{-i H t / \hbar}\ket{\psi_0} = \int dE \ket{\phi_E}\bra{\phi_E}\ket{\psi_0} e^{-i E t / \hbar}.
    \label{eq:timeevolvedstate}
\end{equation}
Taking the overlap of Eq.~\ref{eq:timeevolvedstate} with $\ket{\psi_0}$ and recognizing the initial density of states as $\rho(E) = |\bra{\phi_E}\ket{\psi_0}|^2$, the survival amplitude is the Fourier transform
\begin{equation}
    A(t) = \int_{-\infty}^\infty dE \rho(E) e^{-i E t / \hbar}.
    \label{eq:FTsurvivalamplitude}
\end{equation}
The survival probability is $A^2$.
\begin{figure}[t]
    \centering
    \includegraphics[width = \columnwidth]{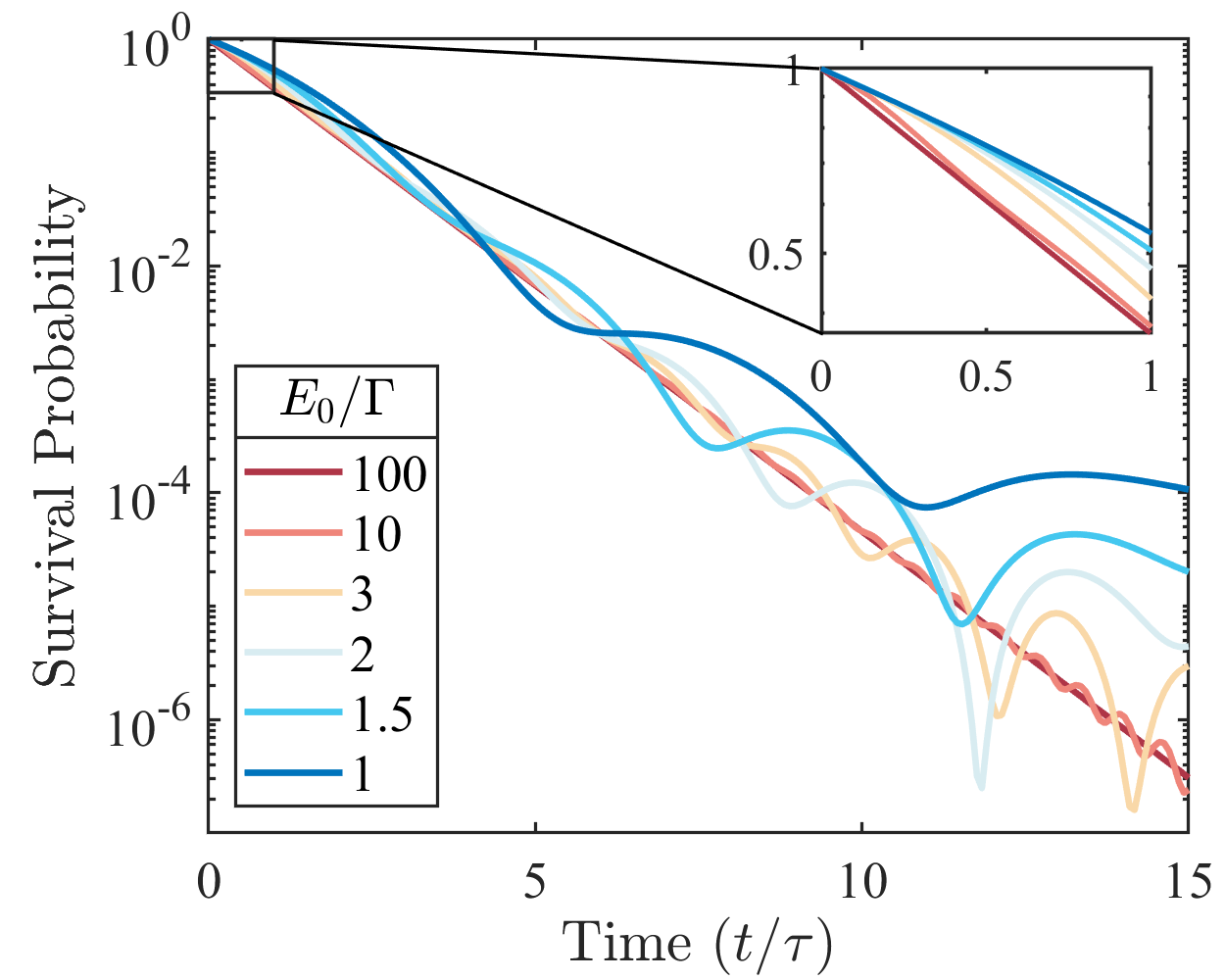}
    \caption{Emergence of non-exponential decay due to truncation of the energy distribution. The survival probability is plotted versus time for various values of $E_0$, as indicated in the legend. The ground state energy is set to 0. $\hbar$ is set to 1 with time measured in lifetimes $\tau$ and energy in linewidths $\Gamma$. The inset highlights the largest deviations in the first lifetime.}
    \label{fig:breit_wigner_ft}
\end{figure}
A simple assumed form for the energy distribution $\rho(E)$ is a Lorentzian or Breit-Wigner distribution: 
\begin{equation}
    \rho(E) = \frac{\Gamma}{2 \pi}\frac{1}{(E - E_0)^2 +  (\frac{\Gamma}{2})^2},
    \label{eq:breitwigner}
\end{equation}
where $E_0$ is the mode and $\Gamma$ is the linewidth. Inserting Eq.~\ref{eq:breitwigner} into Eq.~\ref{eq:FTsurvivalamplitude} and squaring yields the familiar result of exponentially decaying survival probability with decay rate $1/\tau = \Gamma / \hbar$. 

{\em Non}-exponential decay at long times arises from including in this simple argument the fact that real systems necessarily have a lowest energy state, requiring either a truncation of $\rho(E)$ or a bounding of the integral in Eq.~\ref{eq:FTsurvivalamplitude} from below. This alters the form of the survival probability from a pure exponential, giving rise to corrections at long time scales. Fig.~\ref{fig:breit_wigner_ft} shows the non-exponential population dynamics which result from imposing such a lower energy bound.  The absolute square of $A(t)$ is plotted for varying values of the decay product energy $E_0$, demonstrating a clear change from almost purely exponential behavior when $E_0$ is many linewidths away from the ground state to large oscillations and strongly non-exponential dynamics for small values of $E_0$. Here the ground state energy is set to 0.  This slower than exponential decay at very long times is well understood theoretically~\cite{Fonda_1978,Urbanowski2009}, but poses a major challenge for experimental observation due to the small scale of the deviations (note the logarithmic $y$ axis of Fig.~\ref{fig:breit_wigner_ft}) and the many half-lives elapsed before their onset. However, the inset of Fig.~\ref{fig:breit_wigner_ft} reveals that significant non-exponential behavior arises even within the first lifetime when the truncation occurs within a few linewidths of the distribution peak. The scale of these deviations is on the order of $10\%$, which should be readily accessible to detection.

\begin{figure}[t]
    \centering
    \includegraphics[width = \columnwidth]{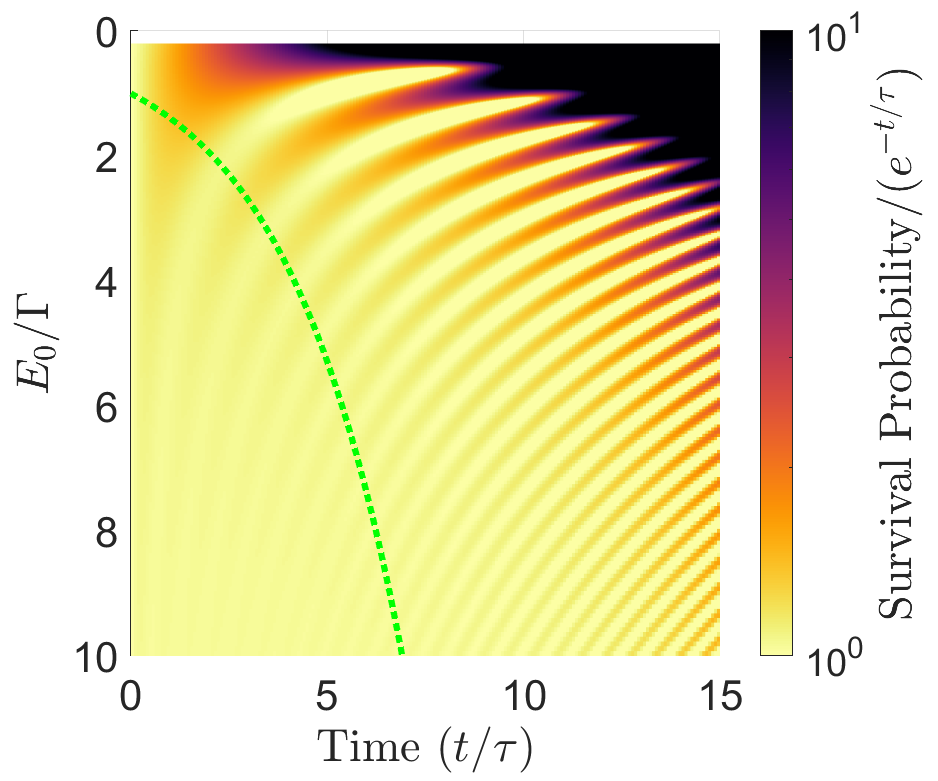}
    \caption{Non-exponential population dynamics as a function of time and the ratio $E_0/\Gamma$. Note that the survival probability color map is normalized to an exponential law in time, with black indicating an order of magnitude population excess with respect to the exponential decay prediction. Dotted green line is the prediction for the onset of non-exponential decay as given by Eq.~\ref{eq:greenlandlimit}.}
    \label{fig:decaymap}
\end{figure}

It is instructive to compare these results to the prediction of Ref.~\cite{Greenland_seeking} that the timescale $\tau_L$ for long time deviations is approximately given by
\begin{equation}
   \tau_L\simeq 3 \tau \log\left(E_0 \tau/ \hbar \right)=3\tau\log\left(E_0 / \Gamma\right) \label{eq:greenlandlimit},
\end{equation}
where $E_0$ is the energy released in the decay. Intuitively, this indicates that $\tau_L / \tau$ (or $E_0 / \Gamma$) cannot be much larger than unity in order for there to be a significant remaining population to exhibit non-exponential behavior. In Fig.~3, we map out the numerical integration of Eq.~\ref{eq:FTsurvivalamplitude} for the range of $E_0/\Gamma = 0.2-10$. We also plot the results of Eq.~\ref{eq:greenlandlimit}. While the prediction is qualitatively correct, for $E_0/\Gamma \approx 2-3$ it somewhat overestimates the onset time; there is clear non-Markovian behavior even within the first time constant. Note the logarithmic scale of the color bar. Overall, though, Fig.~\ref{fig:decaymap} confirms the intuitive result of Eq.~\ref{eq:greenlandlimit} that minimizing the decay product energy with respect to the decay rate yields the largest signal for non-exponential behavior.

In passing, we note that {\em short}-time deviations from exponential decay arise from a related but distinct mechanism: the finite expectation value of energy leading to a survival probability with initially vanishing time derivative~\cite{khalfin_shorttime_kmeson}. This phenomenon underlies the quantum Zeno effect, which was also first realized experimentally with cold atoms~\cite{raizen_zenoantizeno}.

\section{Probing non-exponential decay in modulated optical lattices \label{sec:experiment}}
\begin{figure}[t]
    \centering
    \includegraphics[width=\columnwidth]{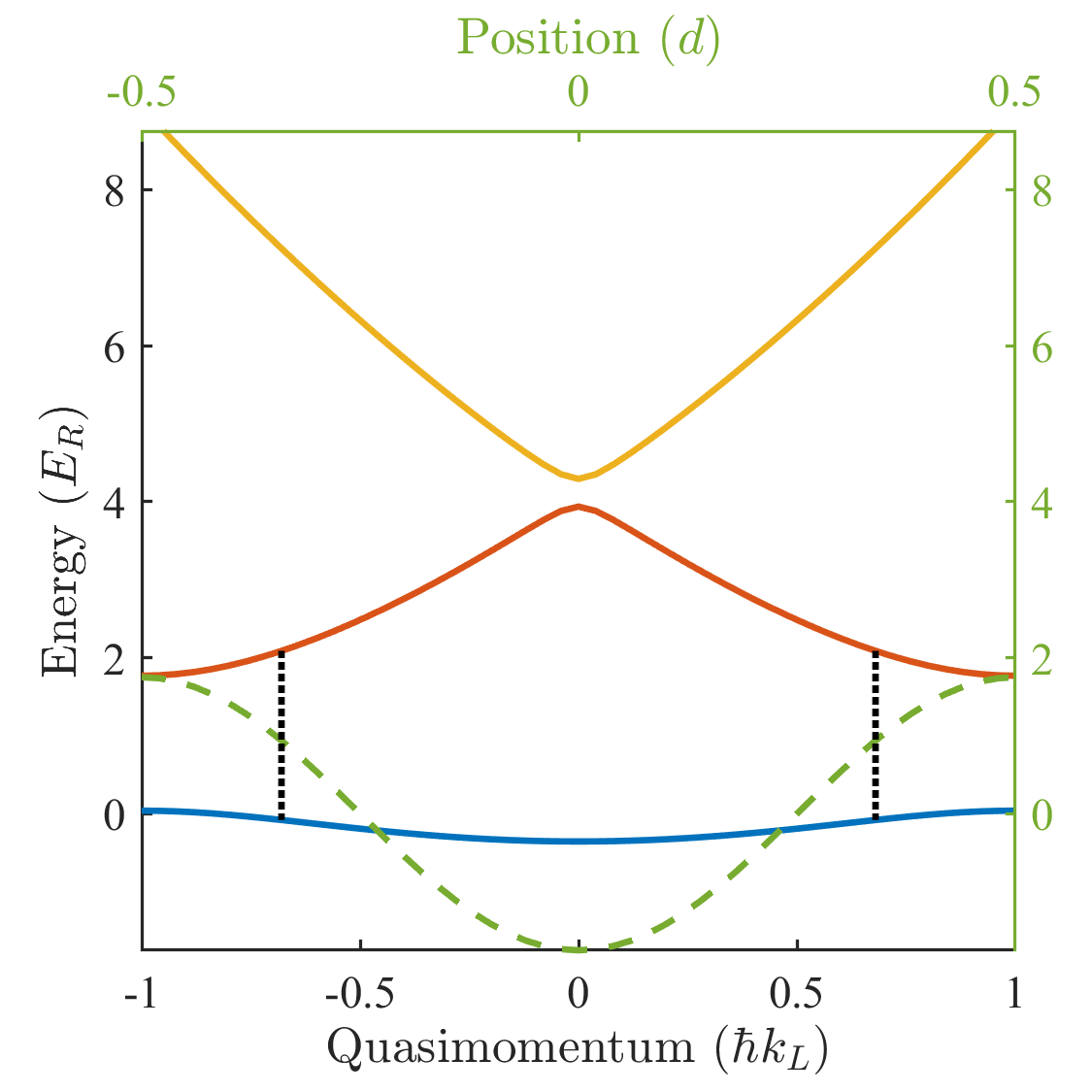}
    \caption{Band structure of a $3.5$ $E_R$ deep undriven optical lattice. Solid lines are the lowest three energy bands. Dashed line overlays the lattice potential in position space (top axis). Dotted black line depicts the drive hybridization scheme used in Figure \ref{fig:tunablegapexp}, ignoring coupling to higher bands.}
    \label{fig:blochbands}
\end{figure}

The experimental probe of non-exponential decay we propose here is based on Bloch oscillations of an ultracold atom ensemble through partially avoided band crossings in modulated optical lattices. Our experimental platform consists of a Bose condensate of $10^5$ $^7$Li atoms in a far-red-detuned ($\lambda = 1064$ nm) optical lattice. Interatomic interactions can be eliminated entirely using the shallow zero-crossing below $^7$Li's broad magnetic Feshbach resonance~\cite{Hulet-tunableinteractionsin7Li}; this crucially allows us to probe the fundamental question of non-exponential decay of a single emitter. The lattice induces an energy band structure, shown in Figure \ref{fig:blochbands}, which can be probed with Bloch oscillations induced by an applied tilt of the harmonic magnetic confinement. In fact the high tunneling rate of $^7$Li enables spatial resolving of different band populations {\em in situ} without the use of band maps or time-of-flight imaging~\cite{RSBO-PRL}. Time-periodic modulation of the lattice depth enables the creation of hybridized Floquet-Bloch bands~\cite{warpdrivePRL} with a drive-dependent band structure; as argued below this is a key capability for realistic observation of non-exponential decay. 

We begin by considering the use of Bloch oscillations in an {\em undriven} lattice as a  probe of decay dynamics. In such an experiment, the atoms are adiabatically loaded into the ground band of the lattice, then undergo Bloch oscillations due to the applied force from the inhomogeneous magnetic potential. Ignoring the field curvature, the main correction to the single-band approximation for the Wannier-Stark problem comes from tunneling between adjacent bands.  As the atoms traverse the edge of the Brillouin zone, they have a chance to ``decay'' by tunneling across the first band gap once per Bloch cycle. The feasibility of observing long-time deviations from exponential decay in such an experiment can be quantitatively estimated using a Landau-Zener model of interband tunneling~\cite{wannierstarkreview}. Semiclassically, the probability of tunneling across the $n$th band gap $\Delta_n$ in a single Bloch cycle is
\begin{align}
    P_n =\exp\left[-\frac{\pi^2}{2}\frac{\Delta_n^2}{h f_B \frac{\partial}{\partial q}|\mathcal{E}_n-\mathcal{E}_{n-1}|}\right],
    \label{eq:lzprob}
\end{align}
where $f_B$ is the Bloch frequency and $\mathcal{E}_n$ is the dispersion of the $n$th band in the free particle limit, indexed with $n=0$ as the ground band. The derivative with respect to the undimensionalized quasimomentum ($q = k/k_L$ and $k_L = 2 \pi/\lambda$) is evaluated at the point of avoided crossing. By modeling the decay as a discrete process happening once per Bloch cycle and then taking a continuum limit, the effective tunneling rate across the $n$th band gap is approximated as
\begin{equation}
    \frac{1}{\tau} \approx f_B \log(\frac{1}{1 - P_n}).
    \label{eq:lzrate}
\end{equation}
In a shallow lattice, tunneling between all excited bands is large and we can treat them as a continuum, so we need only focus on tunneling across the first band gap. In calculating the probability $P_1$ to tunnel out of the ground band, we have $\frac{\partial}{\partial q}|\mathcal{E}_1 - \mathcal{E}_0| = 4$ $E_R$ evaluated at the Brillouin zone edge $q=1$, where the recoil energy is $E_R = \hbar^2 k_L^2 / 2 m$ with $m=7$ amu. Eqs.~\ref{eq:lzprob}~and~\ref{eq:lzrate} reveal two important parameters for optimizing the decay rate of static Bloch oscillations: the band gap $\Delta_1$ and the Bloch frequency $f_B$. These cannot be tuned arbitrarily, though the band gap is minimized for low lattice depths and the Bloch frequency is maximized for large magnetic field gradients. Our experiment can reliably achieve Bloch frequencies $f_B \approx 100$ Hz and minimum usable lattice depths of around $1 E_R$, yielding $\Delta_1 \simeq 0.5$ $E_R$. Inserting these values into Eqs.~\ref{eq:lzprob} and \ref{eq:lzrate} reveals that the resulting tunneling probability will be minimal: $P_1\sim10^{-5}$, leading to a decay time $\tau \sim 10^{3}$~s. Clearly more tunability is needed to reach a regime where the predicted long-time  deviations from exponential decay can be observed.  One route could be to use the much stronger gradients attainable in accelerating lattices, but this intrinsically limits the attainable measurement time as the atoms leave the region of interest.  A more flexible possibility is the use of Floquet engineering to tune the bandgap.

\begin{figure*}
\centering
\includegraphics[width = 0.8\textwidth]{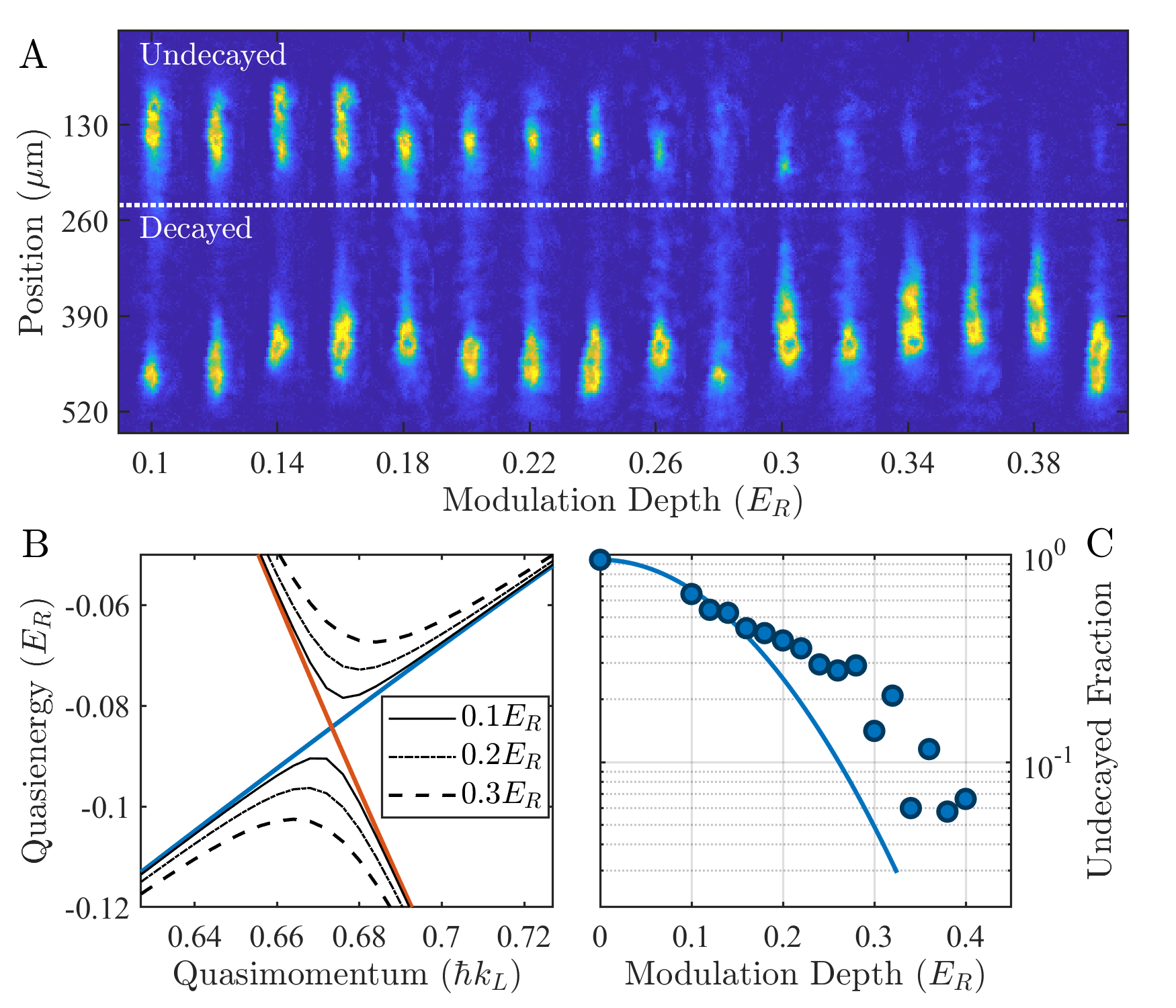}
\caption{Experimental demonstration of Floquet-tunable decay. A) Images of a sample of cold lithium atoms after a single Landau-Zener tunneling event during a Bloch oscillation in a quasienergy band. The "undecayed" upper cloud are those that remain in the ground band of the corresponding undriven system. The lattice depth is $3.5$ $E_R$, the modulation frequency is $55$ kHz, and the Bloch frequency is $27.8$ Hz. B) Calculated quasienergy band structure around the avoided crossing for different modulation depths (indicated in legend). Note the drive-tunable gap. C) Undriven ground band fraction as a function of drive strength. Solid theory line is calculated from Eq.~\ref{eq:lzprob}.}
\label{fig:tunablegapexp}
\end{figure*}

Thus motivated, we consider the addition of time-periodic lattice depth modulation to the experimental protocol outlined above. Resonant coupling of two static bands by such a modulation generically creates a hybrid quasienergy band structure featuring at least one new gap, of a size determined by drive strength rather than lattice depth~\cite{warpdrivePRL}. Fig.~\ref{fig:tunablegapexp}B shows calculated quasienergy band structure near such a gap, for several different values of the drive strength. Tunneling across this tunable gap during a Bloch oscillation in a modulated lattice can realize a much more controllable decay process, in which the decay time can be tuned independently of lattice depth and potential tilt.

To demonstrate this central element of the proposed realization of non-exponential decay, we have experimentally measured tunable Landau-Zener decay in a Floquet-engineered quasienergy band structure. Fig.~\ref{fig:tunablegapexp} presents an experimental measurement of the Landau-Zener decay probability of Eq.~\ref{eq:lzprob} across a Floquet-tunable band gap as a function of drive strength, for the case of resonant driving between the lowest two energy bands. Images of the two spatially resolved band populations after half a Bloch period in the amplitude modulated lattice are shown in panel \ref{fig:tunablegapexp}A, and the calculated band crossing in the quasienergy picture is shown in panel \ref{fig:tunablegapexp}B. The spatial separation between ``decayed'' and ``undecayed'' populations is a consequence of position-space Bloch oscillations in the two different band dispersions~\cite{RSBO-PRL}. Plotting the fraction of undecayed atoms that remain in the ground band, we measure a tunable decay in qualitative agreement with the Landau-Zener tunneling theory of Eq.~\ref{eq:lzprob}, as shown in panel \ref{fig:tunablegapexp}B. Deviations of the data from theory may be the result of uncertainty in the lattice depth or inhomogeneity of the force. Note that in this case, it is actually the atoms that {\em fail} to undergo the tunneling event which correspond to the decayed population. To obtain a decay rate then, we must actually subtract Eq.~\ref{eq:lzprob} from 1. In any case, these results demonstrate the capacity to use lattice modulation to tune the tunneling probability over a wide range, including an enhancement of roughly four orders of magnitude over the tunneling probability in a static band for equivalent conditions. Crucially, this allows $\Gamma$ to approach our achievable Bloch frequencies of up to $100$ Hz, allowing for reasonable experimental run times and detectable non-exponential dynamics. 

\section{Conclusion}

We have proposed a measurement of non-exponential decay of individual emitters which is based on interband tunneling of cold atoms during a Bloch oscillation in a Floquet-engineered quasienergy band. A simple theoretical treatment of expected dynamics indicates that deviations from exponential decay should be measurable. Preliminary experimental tests of the proposed tunable decay mechanism demonstrate widely tunable decay rates and the feasibility of the underlying concept.  These results lay the groundwork for realizing a new experimental probe of universal non-Markovian evolution, and open up new possibilities for exerting quantum control over an irreducible element of non-equilibrium quantum dynamics.

\section*{Acknowledgements}
We thank Andrey Kolovsky, Toshi Shimasaki, and Peter Dotti for useful discussions, and Jeremy Tanlimco and Jared Pagett for experimental assistance. EL acknowledges support from the Knut and Alice Wallenberg Foundation and from the Swedish Research Council (2016-03789). DW is grateful for hospitality provided by the guest researcher program of the Wallenberg Centre for Quantum Technology, and acknowledges support from the National Science Foundation (CAREER 1555313), the Army Research
Office (PECASE W911NF1410154 and MURI
W911NF1710323), and the University of California's Multicampus Research Programs and Initiatives (MRP19-601445). DW and RS acknowledge support from the UCSB NSF Quantum Foundry through Q-AMASE-i program award DMR-1906325.


\pagebreak
\pagestyle{empty}

\end{document}